\documentclass[preprint2]{aastex}

\usepackage{natbib}
\usepackage{graphicx}
\usepackage{pifont}
\usepackage{booktabs}
\usepackage{ragged2e}
\usepackage{mathtools}
\newcommand{\myemail}{gabor.orosz@gmail.com\\Accepted to The Astronomical Journal (January 17, 2017)}

\shorttitle{1612 MHz stellar OH maser astrometry. I. Annual parallaxes}
\shortauthors{Orosz et al.}

\begin{document}

\slugcomment{Accepted to The Astronomical Journal (January 17, 2017)}

\title{Astrometry of OH/IR stars using 1612 MHz hydroxyl masers. \\ I. Annual parallaxes of WX~Psc and OH138.0+7.2}

\author{G.~Orosz\altaffilmark{1}, 
H.~Imai\altaffilmark{1}, 
R.~Dodson\altaffilmark{2}, 
M.~J.~Rioja\altaffilmark{2,3,4},
S.~Frey\altaffilmark{5},
R.~A.~Burns\altaffilmark{1,6}, 
S.~Etoka\altaffilmark{7},
A.~Nakagawa\altaffilmark{1}, 
H.~Nakanishi\altaffilmark{1,8,9}, 
Y.~Asaki\altaffilmark{10,11},
S.~R.~Goldman\altaffilmark{12}, 
and D.~Tafoya\altaffilmark{13}} 

\affil{\altaffilmark{1}Department of Physics and Astronomy, Graduate School of Science and Engineering, Kagoshima University, 1--21--35 Korimoto, Kagoshima 890--0065, Japan}
\affil{\altaffilmark{2}International Centre for Radio Astronomy Research, M468, The University of Western Australia, 35 Stirling Hwy, Crawley, Western Australia 6009, Australia}
\affil{\altaffilmark{3}CSIRO Astronomy and Space Science, 26 Dick Perry Ave, Kensington, Western Australia 6151, Australia}
\affil{\altaffilmark{4}Observatorio Astron\'{o}mico Nacional (IGN), Alfonso XII, 3 y 5, E-28014 Madrid, Spain}
\affil{\altaffilmark{5}Konkoly Observatory, MTA Research Centre for Astronomy and Earth Sciences, Konkoly Thege Mikl\'os \'ut 15--17, 1121 Budapest}
\affil{\altaffilmark{6}Joint Institute for VLBI ERIC, Postbus 2, 7990 AA Dwingeloo, The Netherlands}
\affil{\altaffilmark{7}Hamburger Sternwarte, Gojenbergsweg 112, 21029 Hamburg, Germany}
\affil{\altaffilmark{8}Institute of Space and Astronautical Science, Japan Aerospace Exploring Agency, 3--1--1 Yoshinodai, Sagamihara, Kanagawa 252--5210, Japan}
\affil{\altaffilmark{9}SKA Organization, Jodrell Bank Observatory, Lower Withington, Macclesfield, Cheshire SK11 9DL, UK}
\affil{\altaffilmark{10}Chile Observatory, National Astronomical Observatory of Japan, National Institute of Natural Science, Joaquin Montero 3000 Oficina 702, Vitacura, Santiago, C.P.7630409, Chile}
\affil{\altaffilmark{11}Joint ALMA Observatory, Alonso de Cordova 3107, Vitacura 763 0355, Santiago, Chile}
\affil{\altaffilmark{12}Astrophysics Group, Lennard-Jones Laboratories, Keele University, Keele, Staffordshire ST5 5BG, UK}
\affil{\altaffilmark{13}Chalmers University of Technology, Onsala Space Observatory SE--439 92 Onsala, Sweden}

\email{\myemail}

\begin{abstract}
We report on the measurement of the trigonometric parallaxes of 1612~MHz hydroxyl masers around two asymptotic giant branch stars, WX~Psc and OH~138.0+7.2, using the NRAO Very Long Baseline Array with in-beam phase referencing calibration. 
We obtained a 3$\sigma$ upper limit of $\le$5.3~mas on the parallax of WX~Psc, corresponding to a lower limit distance estimate of $\gtrsim$190~pc.
The obtained parallax of OH~138.0+7.2 is 0.52~$\pm$~0.09~mas ($\pm$18\%), corresponding to a distance of 1.9~$^{+0.4}_{-0.3}$~kpc, making this the first hydroxyl maser parallax below one milliarcsecond. We also introduce a new method of error analysis for detecting systematic errors in the astrometry. Finally, we compare our trigonometric distances to published phase-lag distances toward these stars and find a good agreement between the two methods.
\end{abstract}


\section{Introduction}
\label{sec:intro}

Relative astrometry using very long baseline interferometry (VLBI) has proven to be very successful for conducting trigonometric parallax measurements of maser (and also radio continuum) sources. Masers are excellent tracers of various environments related to the young and evolved stellar populations in the Milky Way Galaxy: high-mass star forming regions (HMSFRs), asymptotic giant branch stars (AGBs) or their massive counterparts, red supergiants (RSGs). CH$_{3}$OH or H$_{2}$O masers at the relatively high radio frequencies of 12 and 22~GHz respectively are excellent for high-precision parallax measurements of HMSFRs at a $\sim$10~$\mu$as-level. This astrometric precision is achieved by using techniques that enable us to calibrate errors due to atmospheric effects. The geodetic blocks in VLBI observations \citep{reid2009} and the dual-beam system \citep{honma2008} introduced in the Japanese VLBI Exploration of Radio Astrometry (VERA) are examples of such calibration techniques, both designed to address the dominant error sources above $\sim$10~GHz: the static (temporally stable and systematic) and dynamic (short-term turbulent and random) terms of the non-dispersive excess path delays caused by the troposphere, respectively.

Thanks to these calibration techniques, there have been a series of successful parallax measurements of AGB stars using mostly circumstellar H$_{2}$O masers \citep[see][and references therein]{nakagawa2016}, with also a few results from SiO masers at 43~GHz \citep[e.g.][]{min2014}. H$_{2}$O masers have also been used to measure distances to a couple of post-AGB stars \citep{tafoya2011,imai2011,imai2013b} and RSGs \citep[e.g.][]{asaki2010}. However, H$_{2}$O masers are neither the strongest nor the most stable of the stellar masers for measuring the parallaxes of AGB stars. Instead, the strongest and most commonly found ones are the 1.6~GHz ground state OH masers, with thousands of known sources in the Milky Way Galaxy \citep{engels2015a}.

In the case of OH/IR stars -- AGB stars heavily enshrouded in optically thick circumstellar envelopes (CSEs) -- OH masers are suitable for astrometry, especially the strongest line at 1612 MHz \citep[e.g.][]{herman1985}. These masers are located in the CSE at a distance of several hundred stellar radii from the central host star, expanding outward at terminal velocity. OH masers are pumped by infrared continuum background radiation to which stellar photons are converted by a heavy dust shell. As a result, 1612~MHz masers are excellent tracers of OH/IR stars: they are saturated, radially amplified and located in relatively calm regions, with strong features stable over decadal timescales \citep[e.g.][]{etoka2000}.

However, VLBI astrometric observations at low frequencies (especially below approximately 2~GHz, i.e. \hbox{L-band}) are challenging to calibrate accurately due to the dominant ionospheric error contributions with typical residual path length errors of hundreds of centimeters at 1.6~GHz, compared to only a few centimeters from the troposphere. These dispersive terms are slowly changing spatial irregularities of plasma density in the atmosphere that cause serious direction dependent errors in astrometry. In turn, they drastically degrade the accuracy of conventional phase-referencing techniques by introducing systematic astrometric offsets into the observations.

Despite the challenges, there have been a few results of annual parallax measurements using OH masers. \citet{vanlangevelde2000} and \citet{vlemmings2003} used the two main-line OH masers at 1665 and 1667 MHz with mostly a conventional source-switching phase referencing strategy to measure parallaxes at a $\sim$1~mas-level precision. \citet{vlemmings2007} refined and continued these measurements, using in-beam phase referencing -- i.e. simultaneously observing the maser with a reference calibrator which lies within the same antenna beam  -- to push the astrometric precision firmly into the sub-mas regime. As we will also discuss in this paper, the problems encountered at low frequencies can be mitigated by substantially decreasing the target--calibrator separations, from a few degrees down to a few tens of arcminutes. 

Looking beyond spectral line VLBI, we see that L-band pulsar astrometry has flourished in the past decade, as those observations are not hindered by resolved spatial and velocity structures like OH masers. Since pulsars are continuum sources, wider bands, higher recording rates and pulsar gating -- recording signals only during on-pulse periods -- can be employed to significantly increase signal-to-noise ratios and, as a result, reduce (random noise-induced) astrometric errors. For the systematic errors, so far, there have been two main ionospheric calibration strategies for L-band continuum astrometry: measure and remove the dispersive component of the ionospheric delays by using wide-spread bands to detect its frequency-dependent curvature \citep{brisken2000,brisken2002}; or use in-beam astrometry to almost completely remove dynamic (random) and mitigate static (systematic) error terms \citep{chatterjee2001,chatterjee2009}. \citet{deller2013,deller2016} demonstrated that by using in-beam calibrators with careful scheduling and data reduction, it is possible to measure parallaxes at a $\sim$10~$\mu$as-level precision despite the low frequency.

An important side note is that ionospheric errors are not only a problem for L-band astrometry, but can dominate the error budget even at 5~GHz \citep[see e.g.][]{krishnan2015,kirsten2015}. In addition, in-beam astrometry can be limited by the availability of suitable calibrators, and by residual systematic errors in the measurements due to non-zero source separations. We are therefore working on developing an alternative multi-calibrator approach \citep[][]{rioja2009,dodson2013} that can fully remove even the effects of the static ionosphere and make our astrometry completely free of systematic errors. This technique, known as MultiView, will be demonstrated in a forthcoming paper \citep{rioja2017}; while our current focus centers on exploring the limits of in-beam maser astrometry at low frequencies. 

In this paper, we report on the VLBI observations of 1612~MHz OH masers to measure the trigonometric parallaxes and proper motions of two long-period variable OH/IR stars, WX~Psc and OH~138.0+7.2. Their respective pulsation periods are 650 and 1410 days \citep{engels2015b}, and both exhibit high mass loss rates of $\sim$10$^{-5}$~M$_\odot$~yr$^{-1}$ \citep[calculated based on the method in][]{goldman2017}. We describe the VLBI observations in Sect.~\ref{sec:obs}, then the flow of data reduction and maser detections in Sect.~\ref{sec:metho}. A comprehensive astrometric error analysis is conducted in Sect.~\ref{sec:errors}, describing the different error terms and their effects on our measurements. We also present a new method to try to identify systematic errors and estimate the uncertainties in our astrometric measurements. The astrometric results and parallaxes are presented in Sect.~\ref{sec:results}, along with a comparison to distances derived by the phase-lag technique \citep[e.g.][]{vanlangevelde1990}. Section~\ref{sec:conclusions} gives a summary and outlook to a follow-up paper that will introduce theoretical topics attainable with our present and future OH maser parallaxes.

\section{Observations}
\label{sec:obs}

We observed two OH/IR stars, WX~Psc (also known as IRC +10011 or OH~128.6$-$50.1) and OH~138.0+7.2 (hereafter abbreviated to OH138) with the NRAO\footnote{The National Radio Astronomy Observatory is a facility of the National Science Foundation operated under cooperative agreement by Associated Universities, Inc.} Very Long Baseline Array (VLBA). Table~\ref{table:sources} lists basic details on the target maser sources and calibrators used. The target sources were selected from the ``Nan\c{c}ay 1612 MHz monitoring of OH/IR stars'' project\footnote{Project home: www.hs.uni-hamburg.de/nrt-monitoring} \citep{engels2015b}, based on having calibrators in the same VLBA beam with precise and reliable positions\footnote{Calibrators with status ``C'' in the Astrogeo Catalog (astrogeo.org).}, i.e. within 0$\fdg$7 of the OH maser sources in our present case. Also, only double-peaked OH sources were considered, which already had their periods and phase-lag distances measured.

\begin{table*}[htbp]
\centering
\caption[]{\label{table:sources}List of observed sources.}
\begin{tabular}{c @{\hskip 6pt} c @{\hskip 6pt} c @{\hskip 6pt} c @{\hskip 6pt} c @{\hskip 6pt} c @{\hskip 6pt} c c @{\hskip 6pt} c}
\tableline
\tableline
\noalign{\smallskip}
\multicolumn{1}{c}{Target} & Period & V$_{\rm LSR}$ & Calibrator & ID & Right ascension & Declination & Sep. & S$_{\rm 1.6GHz}$ \\
\noalign{\smallskip}
& (days) & (km s$^{-1}$) & (J2000) & & ( h \enspace m \enspace s ) & ( \degr \enspace \arcmin \enspace \arcsec ) & (\degr) & (mJy beam$^{-1}$)\\
\tableline
\noalign{\smallskip}
WX~Psc & 650    & 8.9 $\pm$ 0.1   & J0106+1300 & C$_{\rm ib}$        & 01 06 33.35651  & 13 00 02.6039  & 0.40 & 70\\
                               &            &                & J0121+1149 & C$_{\rm 1}$/FF & 01 21 41.59504  & 11 49 50.4130  & 3.81 & 2100\\
\noalign{\smallskip}
OH138.0+7.2       & 1410 & $-$37.7 $\pm$ 0.1  & J0322+6610 & C$_{\rm ib}$       & 03 22 27.22883  &  66 10 28.3005  & 0.70 & 750\\
                               &            &                & J0257+6556 & C$_{\rm 1}$      & 02 57 01.34302  &  65 56 35.4270  & 2.92 & 270\\
                               &            &                & J0102+5824 & FF       & 01 02 45.76238  &  58 24 11.1366  & 18.74 & 1300\\
\noalign{\smallskip}
\tableline
\noalign{\smallskip}
\end{tabular}\\
\vspace{-10 pt}
\justify
{\bf Notes.} For the stellar coordinates refer to Sect.~\ref{sec:metho}, while for the precise OH maser offsets to Table~\ref{table:astrometry}. Stellar pulsation periods are from the ``Nan\c{c}ay 1612 MHz monitoring of OH/IR stars'' project. Calibrators were selected using the Astrogeo Catalog and their coordinates are accurate to $\lesssim$0.3 mas. Systemic velocities and $S_{\rm 1.6GHz}$ flux densities are measured using the VLBA observations. Calibrator IDs refer to Fig.~\ref{fig:obs-setup} and go as follows: C$_{\rm ib}$= in-beam; C$_{\rm 1}$=secondary; FF=fringe finder.
\end{table*}

\begin{figure*}[htbp]
\centering
\includegraphics[width=1\textwidth, trim=0mm 13mm 0mm 13mm, clip]{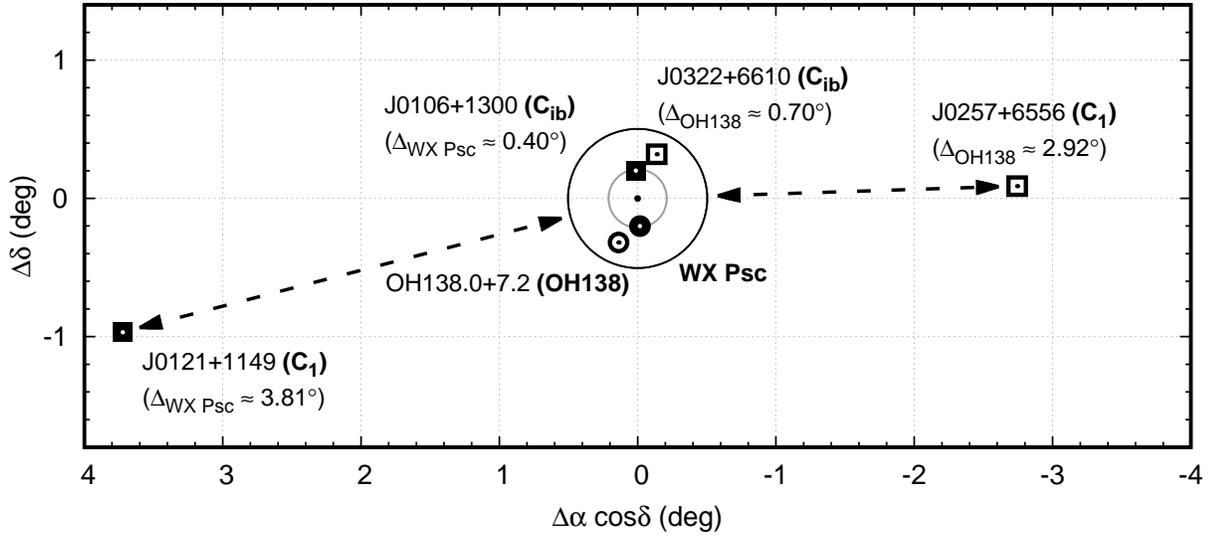}
\caption{Observing setup of the astrometric measurements for both targets plotted together. Solid and open symbols represent the sources associated with the two different monitoring campaigns of WX Psc and OH138, respectively. The target masers are marked as circles and the calibrators as squares. Angles in parentheses show the separation between the respective targets and calibrators. The two concentric circles represent the half-power beamwidth and full beamwidth of the VLBA. Dashed arrows denote source switching (with 5--15 min cycles). Coordinates are relative to the target pointing centers, different for each campaign.}
\label{fig:obs-setup}
\end{figure*}

Each source was observed at four epochs over a period of one year. The details of the observing sessions are given in Table~\ref{table:observations}, whereas the source scan pattern on the sky is shown in Fig.~\ref{fig:obs-setup}. The sessions were scheduled near the peaks of the sinusoidal parallax signatures to maximize the sensitivity for parallax detection and ensure that we can separate the linear proper motions from the parallactic modulations. The desired sampling was only partially achieved due to failed observing epochs and other technical difficulties. All our observations are publicly available from the NRAO VLBA archive under project codes BO047A and BO047B.

\begin{table*}[htbp]
\centering
\caption[]{\label{table:observations}Summary of observing sessions conducted with the VLBA.}
\begin{tabular}{c @{\hskip 6pt} c @{\hskip 6pt} r c c @{\hskip 8pt} c @{\hskip 8pt} c @{\hskip 8pt} c c}
\tableline
\tableline
\noalign{\smallskip}
\multicolumn{1}{c}{Target} & Epoch & \multicolumn{1}{c}{Date (DOY)} & MJD & UT range & Project code & Flagged data & Remarks\\
\noalign{\smallskip}
& & & (days) & (day/hh:mm) & & &\\
\noalign{\smallskip}
\tableline
\noalign{\smallskip}
WX~Psc & I & 2014 Aug 01 (213)  & 56870 & 0/09:32$-$0/13:33 & BO047A6 & LA & no FD\\
              & II & 2015 Feb 17 (048)  & 57070 & 0/20:22$-$1/00:22 & BO047A3 & MK, SC &\\
              & III & 2015 Jun 08 (159)  & 57181 & 0/12:06$-$0/16:06 & BO047A7 & MK, SC &\\
              & IV & 2015 Jul 07  (188)   & 57210 & 0/11:12$-$0/15:12 & BO047A4 & MK, SC &\\
\noalign{\smallskip}
OH138.0+7.2 & I & 2014 Feb 16 (047)  & 56704 & 0/22:25$-$1/02:26  & BO047B1 & MK &\\
              & II & 2014 May 07 (127) & 56784 & 0/18:40$-$0/22:40  & BO047B2 & KP, PT & no NL\\
              & III & 2014 Aug 07 (219)  & 56876 & 0/11:08$-$0/15:08 & BO047B3 & KP &\\
              & IV & 2015 Feb 22 (053)  & 57075 & 0/22:02$-$1/02:02  & BO047B4 & SC &\\
\noalign{\smallskip}
\tableline
\noalign{\smallskip}
\end{tabular}\\
\vspace{-10 pt}
\justify
{\bf Notes.} Flags refer to data from antennas not used in the astrometry, which were identified using methods described in Sect.~\ref{subsec:triangle}. In the NRAO archive WX~Psc has several failed epochs due to recording or correlation problems: BO047A1, A5, \& A7.
\\{\bf VLBA stations.} BR=Brewster, WA; FD=Fort Davis, TX; HN=Hancock, NH; KP=Kitt Peak, AZ; LA=Los Alamos, NM; MK=Mauna Kea, HI; NL=North Liberty, IA; OV=Owens Valley, CA; PT=Pie Town, NM; SC=Saint Croix, VI
\end{table*}

Each observing session was 4\,h long to ensure sufficient $(u,v)$ coverage and sensitivity, spending $\sim$70\% of the time on the in-beam calibrator (C$_{\rm ib}$) and OH maser pairs with intermittent observations of fringe finders (FF) in every 2 hours and bright secondary calibrators (C$_{\rm 1}$) in every 15 min. For the target scans, antennas were pointed halfway between the OH maser and C$_{\rm ib}$ positions, except for Epochs I--II of the WX~Psc sessions, where the pointing center was the OH maser source. The 2-bit quantized signals were recorded on Mark5C units in dual circular polarization with 128 Mbps using 4 intermediate frequency (IF) bands, each with a bandwidth of 4~MHz. The IFs were spread out over 300~MHz, centered around the four ground-state OH maser lines of 1612, 1665, 1667, 1720~MHz and the H\,{\sc I} line at 1420~MHz. Each band had a channel spacing of 1.95~kHz, corresponding to a velocity resolution of 0.36~km~s$^{-1}$. Note that for the last two epochs of WX~Psc (BO047A7, \& A4) we used a slightly modified observing setup for our parallel investigations of high-precision astrometry at L-band \citep[described in][but not related to this work]{rioja2017}. The on-source time on the target was decreased to 40\%; the switching cycle to C$_{\rm 1}$ (and two other calibrators) was decreased to 5~min and the IF bandwidths were increased to 8~MHz (256~Mbps recording rate), but keeping the same spectral resolution. The changed setup did not affect our in-beam parallax measurements.

Since the masers were always observed simultaneously with the C$_{\rm ib}$ phase reference calibrators in the same primary beam, all bands were correlated with VLBA-DiFX \citep{deller2007,deller2011} in a single run using two phase centers set to the maser and C$_{\rm ib}$ calibrator positions. Phase tracking centers for the calibrators were set to the coordinates described in Table~\ref{table:sources}, whereas for the OH maser targets the following \hbox{a-priori} J2000.0 positions were used during correlation: $(\alpha,\delta)_{\text{WX~Psc}}^{\text{a priori}}$=(01$^{\rm h}$06$^{\rm m}$25$\fs$98,$+$12$\degr$35$\arcmin$53$\farcs$0) and $(\alpha,\delta)_{\text{OH138}}^{\text{a priori}}$=(03$^{\rm h}$25$^{\rm m}$08$\fs$80,$+$65$\degr$32$\arcmin$07$\farcs$0). For details on the observations and correlator output, we refer to the VLBA File Server\footnote{VLBA File Server: www.vlba.nrao.edu/astro/VOBS/astronomy}.

\section{Data Reduction and Maser Detections}
\label{sec:metho}

The data analysis was carried out using the NRAO Astronomical Image Processing System (AIPS) package with an in-beam phase referencing strategy. Flux density calibration was performed using system temperatures and gain information recorded at each station. The Earth orientation parameters from the VLBA correlator were refined by the U.S. Naval Observatory final solutions. Initial ionospheric delay corrections were performed using the NASA Jet Propulsion Laboratory IONEX files, which contain zenith total electron content (TEC) maps derived from global navigation satellite system observations. Finally, phases were corrected for parallactic angle effects before phase referencing.

\begin{figure*}[htbp]
\centering
\includegraphics[width=0.49\textwidth, trim=32mm 2mm 32mm 0mm, clip]{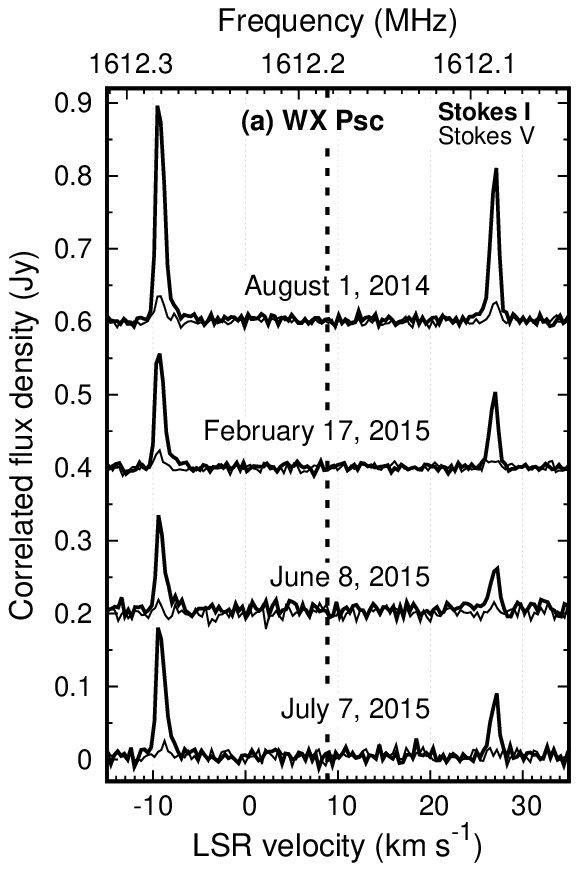}
\includegraphics[width=0.49\textwidth, trim=32mm 2mm 32mm 0mm, clip]{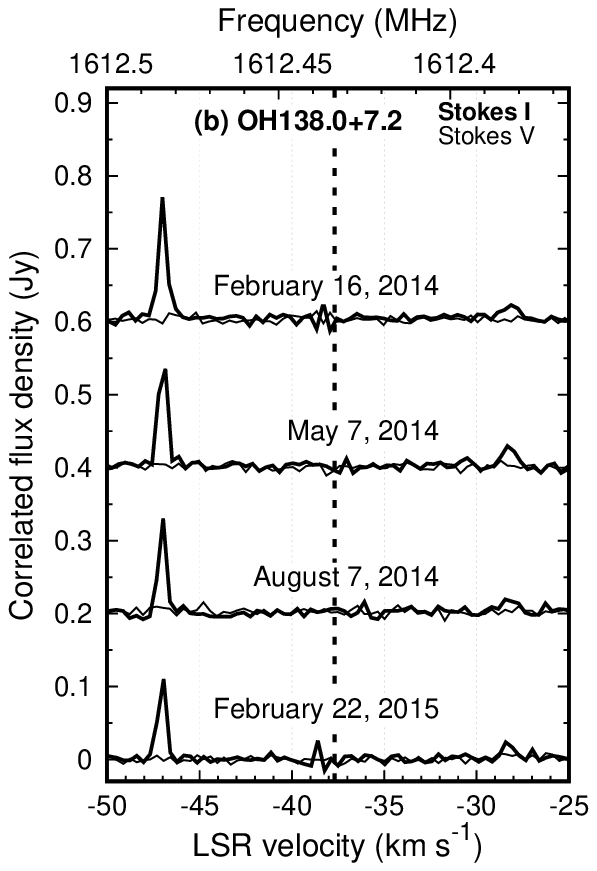}
\caption{
Scalar averaged cross-power spectra of the 1612 MHz OH maser emission for each epoch towards (a) WX~Psc and (b) OH138. Thick and thin lines denote the Stokes I (total intensity) and V (degree of circular polarization) parameters, respectively. The spectra are unsmoothed and have a channel spacing of 1.95 kHz (0.36 km s$^{-1}$). Maser spots used for the parallax fitting are located in the brighter blueshifted peaks. Thick dashed lines mark the systemic velocities, calculated by averaging the strongest blue- and redshifted maser channels from all epochs. Spectra of different epochs have been shifted along the vertical axis for clarity.}
\label{fig:spectrum}
\end{figure*}

We flagged channels contaminated by radio-frequency interference, then performed instrumental phase calibration using single FF calibrator scans with the phase rates zeroed. Next, scans on C$_{\rm 1}$ were used to determine the group delays, phase rates and bandpass characteristics. Fringe-fitting was performed by averaging polarizations, as Stokes V values were always less than $\sim$10\% of Stokes I, making the difference between polarizations negligible (see Fig.~\ref{fig:spectrum}). After applying this calibration to the Maser--C$_{\rm ib}$ pair, the final phase calibration solutions were obtained using C$_{\rm ib}$ (with IFs and polarizations averaged) and transferred to all channels in the maser scans. Thus WX~Psc and OH138 were effectively phase referenced to their respective in-beam calibrators. Finally, phases were rotated to shift the phase-tracking center to the vicinity of the maser emission before imaging. The shifted J2000.0 phase-tracking centers used for astrometry are: $(\alpha,\delta)_{\text{WX~Psc}}^{\text{shifted}}$=(01$^{\rm h}$06$^{\rm m}$26$\fs$02574,$+$12$\degr$35$\arcmin$52$\farcs$8242) and $(\alpha,\delta)_{\text{OH138}}^{\text{shifted}}$=(03$^{\rm h}$25$^{\rm m}$08$\fs$42975,$+$65$\degr$32$\arcmin$07$\farcs$0900).

Though our in-beam phase referencing observations were affected by the primary beam attenuation of the antennas, a correction scheme was not applied in the present observations. As a result, the flux density scales mentioned in this paper are systematically lower than the true values. Using an Airy disk model presented in \citet{middelberg2013} with an antenna diameter of $D$=25.47\,m, we estimate that the amplitudes of WX~Psc and J0106+1300 are lower by $\sim$47\%, and those of OH138 and J0322+6610 are lower by $\sim$90\%. Uncalibrated amplitudes do not seriously affect our astrometric measurements, because the phase pattern in the primary beam is expected not to have a significant systematic offset and also the resulting instrumental error is believed to be constant throughout our observing sessions. However, in order to minimize possible error sources and achieve better signal-to-noise ratios, proper flux density corrections should be applied in future in-beam astrometric observations by adopting a suitable primary beam model and beam squint corrections as done in \citet{middelberg2013}.

We observed all four ground-state OH maser transitions, but could only detect the double-peaked 1612~MHz satellite lines for both OH/IR stars. The full resolution VLBA spectra of the 1612 MHz OH masers are shown in Fig.~\ref{fig:spectrum}. In both profiles, the blueshifted features relative to the stellar systemic velocities have peak flux densities of $\sim$0.2~Jy, compared to $\sim$20~Jy from single dish observations with, e.g, the Nan\c{c}ay Radio Telescope (NRT), meaning we could only recover $\sim$1\% of the total maser emission from the OH regions. The redshifted features have a similar flux density recovery percentage for WX~Psc, but are almost completely missing for OH138, perhaps due to more serious foreground scattering from the circumstellar envelope. However, in all cases maser emission on baselines longer than $\sim$4000~km are resolved out significantly for both the blue- and redshifted regions, similar to that seen in \citet{imai2013a}.

\section{Astrometric Error Analysis}
\label{sec:errors}

\subsection{Theoretical Predictions}
\label{subsec:aris}
The dominant error sources after our in-beam phase referencing are composed of various components of uncompensated atmospheric terms, and additional contributions from source/velocity structures and thermal noise \citep{reid2014}. Since the in-beam calibrators are observed simultaneously in the same beam with the maser targets, the derived phase solutions can be directly applied to the masers and do not have to be interpolated in time between the calibrator scans. This mitigates the effects of temporal phase fluctuations, a dynamic term causing random errors in the astrometry. The small target--calibrator separations also reduce excess path errors.

Given our observing parameters (1.6 GHz, 0 min switching time due to simultaneous observations, 0$\fdg$4$-$0$\fdg$7 target--calibrator separations) we can estimate the expected errors from the static and dynamic components of the troposphere and ionosphere \citep{asaki2007}, by assuming a typical zenith path error of 3~cm \citep{reid1999}, a 6~TECU uncertainty\footnote{TEC Unit; 1~TECU = 10$^{16}$ electrons m$^{-2}$} in the adopted ionospheric maps \citep{ho1997}, and a typical zenith angle of 45$\degr$. This predicts, per baseline, a dynamic ionospheric phase error of 3$\degr$~/~5$\degr$, and a static ionospheric phase error of 16$\degr$~/~28$\degr$ for WX~Psc~/~OH138. Due to the low frequency, the non-dispersive tropospheric phase errors are negligible; less than 1$\degr$ per baseline.

The biggest residual errors in L-band are related to {\it spatial} static terms, direction dependent systematic errors from the inadequate modeling of the ionospheric sky-plane TEC distribution. This means that reducing the target--calibrator separation is of utmost importance in mitigating atmospheric errors in low-frequency VLBI astrometry. Also, even with in-beam phase referencing the dynamic terms from ionospheric phase fluctuations are not zero. This is because the traveling waves causing the temporal disturbances in the ionosphere have spatial scales of hundreds of kilometers, which again reflect as residual errors due to the non-zero target--calibrator separations. Assuming these components are independent and adding them in quadrature, we estimate the total atmospheric phase errors per baseline to be 16$\degr$~/~29$\degr$ for WX~Psc~/~OH138. These $\phi_{\rm atmo}$ phase errors per baseline can be roughly converted to $\sigma_{\rm atmo}$ astrometric errors in the VLBA maps as $\sigma_{\rm atmo} \approx \big(\phi^{[\circ]}_{\rm atmo}/360^{\circ}\big) \cdot \big(\theta/\sqrt{N}\big)$, where $\theta$ is the size of the synthesized beam and $N$ is the number of stations in the array. Using $N$=$8$ and beam sizes described in Sect.~\ref{subsec:parallax}, we estimate the total atmospheric errors in our astrometry to be approximately 0.3~mas~/~0.4~mas for WX~Psc~/~OH138, dominated by the effects of the static ionosphere. In bad ionospheric conditions (i.e. having larger residual TEC values), deviations even as large as $\sim$0.7 mas can be expected.

Contribution from the target source structure to systematic astrometric errors is hard to predict, due to the variable behavior in the spatial and velocity structures of masers, and the possibility of multiple maser spots blending together. Our OH maser spots used for the astrometric measurements seem to show complex spatial structures at lower resolutions when using a $(u,v)$~taper of 13~M$\lambda$ (corresponding to baselines of $\sim$2000~km). However, when using the full resolution of our array, we only detect compact, albeit not completely unresolved emission. We also try to minimize the systematic effects caused by the maser velocity structure by fitting the parallax using individual spots with the same velocity between epochs. Feature fitting is not feasible due to the few detected spots in each feature\footnote{A maser spot refers to an individual maser brightness peak imaged in one spectral channel, and a maser feature refers to a group of spots which are considered to relate to the same physical maser cloud.}. 

Astrometric errors from the image thermal noise are approximated as $\sigma_{\rm therm} \approx 0.5\cdot\big(\theta/{\rm SNR}\big)$, where $\theta$ is the size of the synthesized beam in case the source is unresolved, and SNR is the signal-to-noise ratio in our VLBA maps. The trade-off of limiting ourselves only to the most compact parts of the OH maser emission for our astrometry -- as an effort to reduce systematic effects from source structure -- is a reduced SNR and thus a larger thermal noise component in the total astrometric error budget. Fortunately, the thermal noise is a random error source, so it has a more benign effect on the parallax measurements than leaving possible systematic errors in our datasets. However, this highlights one of the major difficulties in low-frequency astrometry. As $\sigma_{\rm therm}\propto\nu^{-1}$, where $\nu$ is frequency, the intrinsic limit of astrometry is lower than for CH$_{3}$OH or H$_{2}$O masers at higher frequencies.

We quantitatively investigated the astrometric errors not only by the analytical method described above, but also by simulating our 1.6~GHz VLBA observations with ARIS \citep{asaki2007}. We adopted input parameters based on our observation parameters and typical error values in VLBI observations: the target source is a single circular Gaussian component with a full width at half maximum of 10 mas and a maximum peak of 0.4--0.8~Jy in 1.95 kHz bandwidth. The reference sources for WX~Psc and OH138 are J0106+1300 and J0322+6610 with flux densities of 0.07~Jy and 0.75~Jy in 32 MHz bandwidths, respectively. Because the above source strengths were assigned by referring to our data reduction results, the primary beam attenuation in our in-beam phase referencing was not considered. Imaging was conducted without MK and SC as described in Sect.~\ref{subsec:parallax}. We simulated 200 samples and estimated astrometric errors as the position offsets in RA and Dec from the phase tracking centers that contained 67\% of the simulated positions.

The obtained astrometric errors from the simulations are 1.2~/~0.7 mas for WX~Psc~/~OH138 in the case of a 0.6~Jy target Gaussian component, showing good consistency with our observation results (see Sect.~\ref{subsec:triangle}). Our ARIS simulations show that despite the smaller target--calibrator separation, the astrometric errors of WX~Psc can be worse than that of OH138 mainly because the reference source of WX~Psc is weaker. Although transferred errors from calibrator structure can be ruled out as we modeled their structure before phase referencing, the thermal noise in the transferred visibility phase solutions from the weaker J0106+1300 to WX~Psc is higher than in the case of J0322+6610 and OH138. Our simulations also show that the astrometric errors are particularly sensitive to the peak value of the target Gaussian component: the astrometric error for a maximum peak of 0.4~Jy is a factor of $\sim$1.6 worse than for a maximum peak of 0.6~Jy, while the error for a maximum peak of 0.8~Jy is a factor of $\sim$1.3 better than for a maximum peak of 0.6~Jy. These errors can be different from epoch to epoch as the maser source strength varies. 

We then repeated the same simulations with the ionospheric model errors turned off, and obtained astrometric errors of 1.0~/~0.4 mas for WX~Psc~/~OH138 for a maximum peak of 0.6~Jy. Comparing the two sets of simulations the contribution of the ionospheric model error to the astrometric measurements was found to be $\sim$0.6 mas, which is consistent with the previous analytical estimates. We can also see that the limited bandwidth on the calibrators is a major contributor to the total error budget, which can be avoided in future observations by using larger total bandwidths for an increased continuum sensitivity. The sizes of datasets can be kept manageable by using several correlator passes and spectral ``zooming'', i.e. correlating all scans with a coarse resolution on the full bandwidth and the maser scans with a high resolution on a narrow bandwidth containing the spectral features. This feature is routinely available on both DiFX \citep{deller2011} and SFXC \citep{keimpema2015} correlators used at most VLBI arrays.

\subsection{Empirical Errors from Subarray Imaging }
\label{subsec:triangle}

\begin{figure*}[htbp]
\centering
\includegraphics[width=0.8\textwidth, trim=2mm 0mm 26mm 0mm, clip]{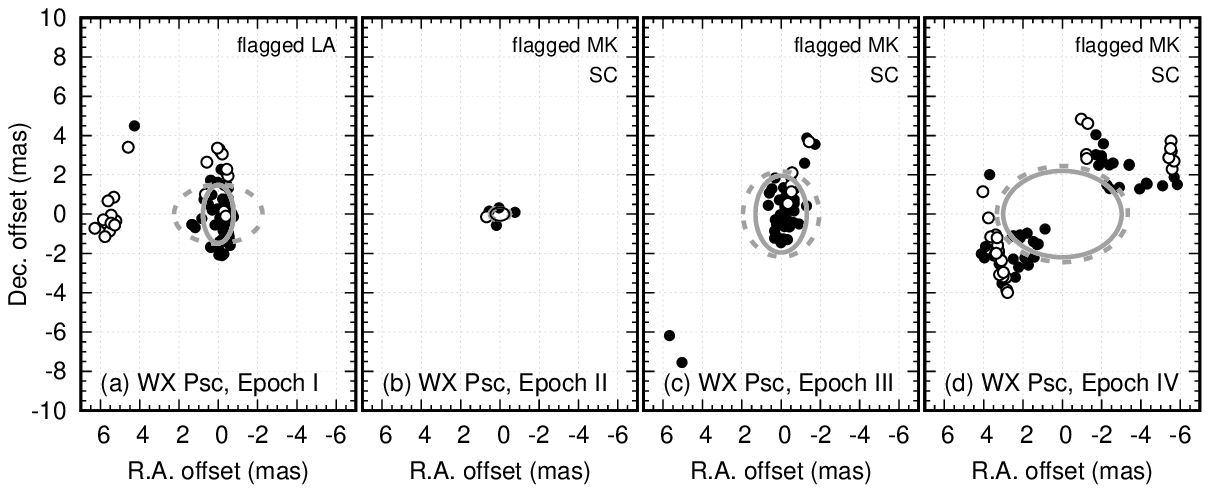}
\includegraphics[width=0.8\textwidth, trim=2mm 0mm 26mm 0mm, clip]{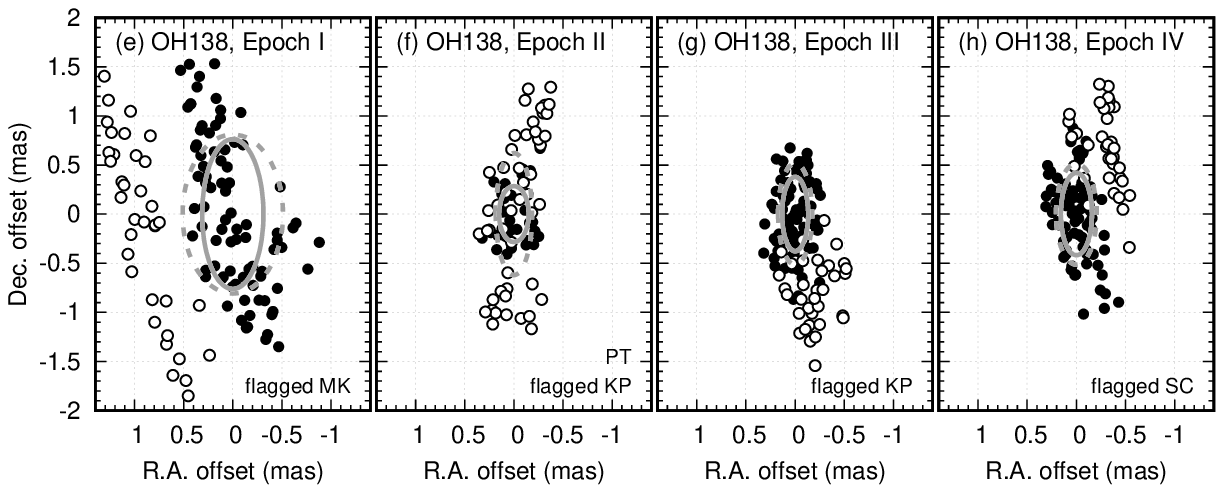}
\caption{
Fitted positions of the brightest maser spot toward (a)--(d) WX~Psc and (e)--(h) OH138 in respective epochs, which are determined using all ($\leq$120) three-antenna subarrays of the VLBA. Open circles mark measurements from triangles that contain antennas flagged in the final astrometric solutions. Offsets are relative to the mean positions derived from the retained data and shifted to zero (filled circles). Dashed and solid ellipses show, respectively, the 1$\sigma$ contour of unweighted Gaussian models fitted to the measured peak positions before and after flagging the indicated antennas. Ellipses have been shifted to the center to make them easier to compare.}
\label{fig:triangles}
\end{figure*}

As we discussed, astrometric errors are composed of systematic and random errors. Because the former can yield systematic shifts in the measured maser positions, their identification is crucial for {\it accurate} astrometry. We do this by imaging the strongest maser channel for each source with all possible three-antenna subarrays of the VLBA using the automated CLEAN procedure in AIPS, then measure the position of the peak in each resulting map with the verb MAXFIT\footnote{For measuring the peak positions, we compared the Gaussian model fitting of IMFIT/JMFIT, the quadratic function fitting of MAXFIT and simply selecting the brightest pixel with IMSTAT. As long as the mapped area was a few times larger than the fitted maser spot and the pixel sizes were adequately small compared to the synthesized beam -- in our case 0.1$\times$0.1 mas -- all three approaches produced nearly identical results.} (see Fig.~\ref{fig:triangles}). By limiting ourselves to three antennas, each with an independent static ionospheric error, we form coherent but shifted images of these subsets of data. Therefore, comparing the maser positions determined from these subarrays clearly expose the antennas that are contaminated by systematic errors, because their subarray images will also be affected and shifted systematically.

Looking at Fig.~\ref{fig:triangles}, it is clear that the subarray imaging worked better for OH138, which has a strong and compact maser spot, and seems less conclusive for WX~Psc where the detected emission is more resolved and much weaker (see Sect.~\ref{subsec:parallax}). In most cases, the distribution of the derived positions is elongated in declination, which is due to the geometry of the VLBA. Among the 120 possible subarrays the majority are dominated by East--West baselines, which yield poorer angular resolution in the North--South direction. This is just a random effect which does not skew the astrometric results (see Fig.~\ref{fig:triangles}a,c,e--h). Next, we discuss the measured patterns, with the specific sessions and epochs shown in bold.

{\bf OH138 session Epochs I--IV.} Systematic shifts could be found and linked to specific antennas (Epoch~I: MK in RA; Epoch~II: PT and KP in positive and negative Dec, respectively; Epoch~III: KP in Dec; Epoch~IV: SC in RA). The size of these shifts are approx. 0.5--1~mas, which agree with our expectations for the static ionospheric error contributions. While substantial offsets found in the case of MK and SC are possible due to the long baselines and difference in antenna elevations, it comes as a surprise that some of the core southwestern antennas (FD, KP, LA, OV, PT) would also be affected by ionospheric model errors to such a degree. From \citet{dodson2016} we found that under normal conditions the typical $\Delta$TEC model differences between VLBA antennas on the mainland US are of the order of 1\,TECU. After the phase referencing corrections with a calibrator $\lesssim$1\degr away, the residual systematic differences would be $\lesssim$1\,TECU$\times$$\sin(1\degr)$$\approx$0.02\,TECU. These would cause systematic phase errors of $\lesssim$6\degr \citep{asaki2007}, which translate to $\lesssim$1\,mas astrometric errors for 600\,km baselines, i.e. the average separation between the core antennas of the VLBA. Although this seems to explain the observed offsets, it is curious that some close-by core stations are more heavily affected than other mainland antennas further away, as would be expected from \citet{dodson2016}. Future observations should be conducted to investigate this issue in more detail. For the present paper, data related to the marked offset antennas were flagged out before making the final image cubes and getting the astrometric solutions used for the parallax fitting (see Fig.~\ref{fig:triangles}e--h).

{\bf WX~Psc session Epoch I.} A systematic shift of $\sim$6~mas can be seen for LA, which is much larger than expected from ionospheric errors. Instead, we suspect that this might be an artifact from the imaging of a weak source with only three antennas. Although we flagged the data related to LA for safety and because flagging provided a slightly better parallax fit, it only changed the parallax value by 4\% which is well below our relative fitting error of $\sim$30\% (see Fig.~\ref{fig:triangles}a).

{\bf WX~Psc session Epoch II.} The maser is too weak to be useful for error analysis as its measured positions show scatter over the whole mapped region of 100$\times$100 mas. As a result, only a few measured positions are found around the mean, with the rest not displayed due to the large scatter (see Fig.~\ref{fig:triangles}b).

{\bf WX~Psc session Epoch III.} Subarray imaging works, but no specific pattern can be seen in the measured maser positions for any antenna. Data points related to antennas MK and SC are randomly scattered over the whole map area and are not displayed beyond the central region (see Fig.~\ref{fig:triangles}c). 

{\bf WX~Psc session Epoch IV.} Measured positions cluster into two complex groups offset by $\sim$6~mas, similar in size to that seen in Epoch I. Also, no antenna could be linked to the pattern. We again suspect that this might be an artifact due to the low image quality of the three-antenna subarrays. The complex geometric structure of the group towards the NW certainly hints at the idea that it might be linked to the different beam patterns of various subarray images. As a result, the measured shifts are not necessarily related to systematic errors and we performed no flagging based on these results (see Fig.~\ref{fig:triangles}d).

To summarize, data flagging for astrometric analysis was done based on the subarray imaging results for all epochs in the OH138 session. All epochs of WX~Psc showed no emission on baselines to MK and SC in their cross-power spectra and measurements of their subarray images showed large scatter, except in Epoch~I. As a result, we flagged out data related to MK and SC in Epochs II--IV, and used the subarray imaging results for flagging Epoch~I.

As a way to evaluate the total astrometric uncertainty of our final measurements, we implement a new technique based on our subarray imaging (see Table~\ref{table:errors}). Calling these new estimates {\it triangle baseline errors}, we calculate them by fitting a 2D unweighted Gaussian model to the distribution of peak positions obtained by the ensemble of subarrays, after flagging the specified antennas (shown as solid grey ellipses in Fig.~\ref{fig:triangles}). There is a potential risk in estimating errors based on a flagging scheme designed to minimize position scatter, as taking out more antennas could result in unreasonably small error values. Keeping this in mind, flagging was kept to the minimum possible, only taking out significantly outlying antennas where the scatter could be explained by a viable error source.

We believe that the resulting values contain all residual error sources and are thus probably less likely to underestimate astrometric uncertainties. However, as noted earlier this method requires the maser spot to be strong enough to image it reliably with only three antennas. We note that using more antennas would decrease the thermal noise and increase the success rate of forming images. However, this would result in less points to analyze and increase the risk of images getting defocused more rather than shifted coherently; although defocusing also affects our three-antenna method due to time averaging over the 4\,h long sessions. This could only be avoided by using short snapshots where the time variable error sources could be considered constant, but the analysis of the resulting images would be hampered by higher side-lobe levels. In any case, using more antennas would make the interpretation of the resulting distribution and identification of possible systematic errors less clear.

\begin{table*}[htbp]
\centering
\caption[]{\label{table:errors}Error estimates of astrometric results}
\begin{tabular}{l @{\hskip 25 pt} c c c c @{\hskip 25 pt} c c c c @{\hskip 25 pt} c c c c}
\tableline
\tableline
\noalign{\smallskip}
Method & \multicolumn{3}{r}{Error in R.A.} & & \multicolumn{3}{r}{Error in Dec.} & & \multicolumn{4}{c}{Total Error}\\
\noalign{\smallskip}
& I & II & III & IV & I & II & III & IV & I & II & III & IV \\
\noalign{\smallskip}
& mas & mas & mas & mas & mas & mas & mas & mas & mas & mas & mas & mas\\
\tableline
\noalign{\smallskip}
& \multicolumn{12}{c}{WX~Psc~~~~Spot \ding{109}}\\
\noalign{\smallskip}
Thermal      & 0.24 &  0.57  & 0.58  & 0.42 & 0.46 & 1.09  & 1.10 & 0.80 &    0.52   &   1.23  &   1.24   &    0.90  \\
Fitting         & 0.71 &  1.00  & 0.75  & 0.67 & 1.04  & 1.50 & 1.39  & 1.04 &    1.26   &  1.80  &  1.58   &    1.24   \\
Triangle      & 0.78 &  0.48$^{\dagger}$  & 1.31  & 3.02$^{\dagger}$ & 1.48  & 0.34$^{\dagger}$ & 1.96  & 2.20$^{\dagger}$ &    1.67   &  0.59$^{\dagger}$  &   2.36   &    3.74$^{\dagger}$  \\
\noalign{\smallskip}
\cmidrule[0.04em](lr){2-13}
& \multicolumn{12}{c}{OH138.0+7.2~~~~Spot \ding{109}} \\
\noalign{\smallskip}
Thermal       & 0.04  & 0.07  & 0.03  & 0.04  & 0.09  & 0.17   & 0.08  & 0.09 &    0.10    &    0.18    &    0.09    &    0.10\\
Fitting          & 0.09  & 0.11  & 0.05  & 0.06   & 0.19  & 0.19  & 0.11  & 0.12 &    0.21    &    0.22    &    0.12    &    0.13\\
Triangle       & 0.31  & 0.15  & 0.13  & 0.16   & 0.76  & 0.29  & 0.38  & 0.42 &    0.82    &    0.33    &    0.40    &    0.45 \\
\noalign{\smallskip}
\tableline
\noalign{\smallskip}
\end{tabular}\\
\vspace{-10 pt}
\justify
{\bf Notes.} I--IV refer to the observing epochs.
Thermal: errors from thermal noise are based on the equation in Sect~\ref{subsec:aris}. Fitting: calculated by least-squares fitting of Gaussian models to the images using the AIPS task IMFIT/JMFIT. The parameters of the image cubes used for the calculations are described in Sect.~\ref{subsec:parallax}. Triangle: errors derived from subarray imaging by fitting Gaussian models to the measured peak positions of all subarrays (see grey solid ellipses in Fig.~\ref{fig:triangles}). All errors are summed in quadrature.\\ $^{\dagger}$ Unreliable triangle baseline errors due to failed subarray imaging (see Sect.~\ref{subsec:triangle}).
\end{table*}

Thermal noise errors are evaluated as the random thermal noise in the maps. Fitting errors are associated with the ability to fit Gaussian models to these maser maps and include the effects of maser source structure. As expected, we find that errors scale as $Thermal < Fitting < Triangle$. In our analyses (see Table~\ref{table:errors}) this trend is evident in all but WX~Psc session Epoch~II, where the subarray imaging failed. The triangle baseline errors of WX~Psc session Epoch~IV also seem problematic and most likely overestimate the real astrometric uncertainties due to the complex structure seen in Fig.~\ref{fig:triangles}d. However, in most cases the values confirm that the triangle baseline errors are the most conservative estimates. They should not be skewed by a high SNR in the image, but contain all possible error sources instead. In conclusion, our final astrometric error estimates are based on these values directly in all but the two problematic epochs. In the case of WX~Psc, triangle baseline errors are 1.1--1.8$\times$ larger than fitting errors where the method works. As a result, for our astrometric analysis we take the final errors as the double of the fitting errors for Epochs II and IV of WX~Psc, although we note that as no real flagging was possible for these epochs (besides MK and SC), the increase in the subarray scatter might be even larger than the values measured using other flagged epochs.

\section{Astrometric Results and Discussion}
\label{sec:results}

\subsection{Maser Maps and Parallaxes}
\label{subsec:parallax}

\begin{figure}[htbp]
\centering
\includegraphics[width=0.45\textwidth, trim=0mm 18mm 0mm 0mm, clip]{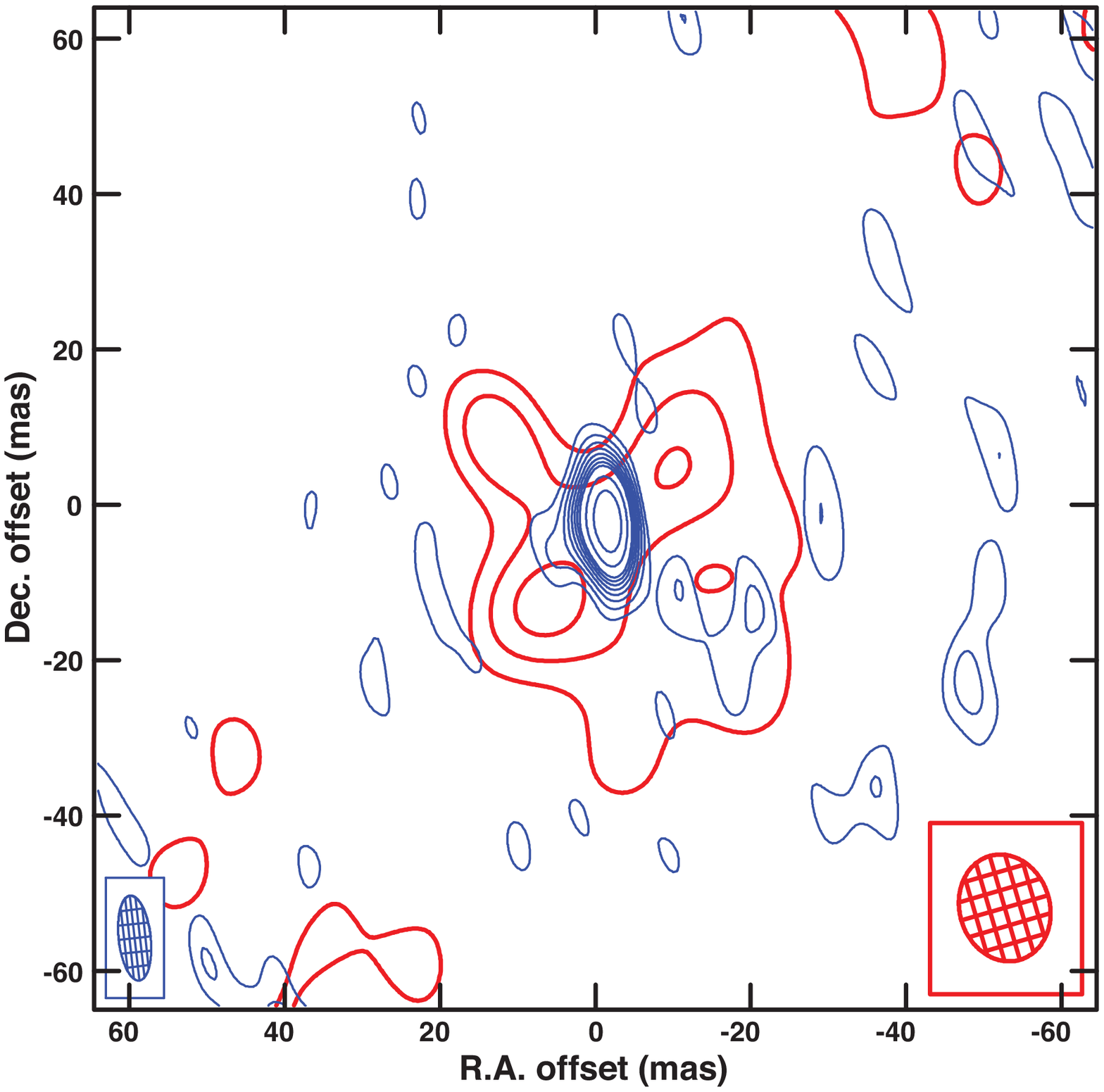}
\caption{
Contour map of 1612 MHz OH maser spots towards OH138 in Epoch III. Blue thin contours show the compact blushifted maser spot used for astrometry at $-$47.0 km s$^{-1}$, while red thick contours show the brightest redshifted maser channel at $-$28.5 km s$^{-1}$. To emphasize faint structures, contour levels run linearly from 3$\times$RMS for 8 steps, then exponentially with a base of $\sqrt{2}$. RMS values refer to the average noise in line-free channels of the image cube. Synthesized beams used for restoring the maps are shown in the bottom corners, with sizes of 13$\times$5~mas (blue) and 16$\times$13~mas (red). The map covers an area of 128$\times$128~mas, centered on the shifted phase tracking center.}
\label{fig:maps}
\end{figure}

Final maser image cubes for proper motion and parallax fitting were produced with the AIPS CLEAN procedure after flagging data listed in Table~\ref{table:observations}, based on our method described in Sect.~\ref{subsec:triangle}. We used uniform weighting for mapping, as that gave the best astrometric results after considering the trade-off between angular resolution and SNR. We detected maser emission over several channels in all epochs of both stars and both peaks, however the redshifted features were always too defocused for astrometry. Typical synthesized beam sizes achieved in the final image cubes were approximately 18~$\times$~8~mas for WX~Psc, and 13~$\times$~5~mas for OH138 due to less flagging. Figure~\ref{fig:maps} shows a typical example from the produced image cubes, which plots the brightest maser channel in both the blue- and redshifted peaks of OH138. 

We detect only a single feature in the blueshifted peaks of WX~Psc and OH138, but they are reasonably compact and stable in both cases to be used for astrometric analysis. We fit 2D Gaussian models to the CLEAN maps of the blueshifted maser emission to determine the astrometric position in each epoch. Faint and extended emission from the brightest redshifted maser spots can only be detected on baselines of $\lesssim$2000~km, as they are completely resolved with smaller synthesized beams. Red- and blueshifted spots clearly overlay for OH138, indicating that we are truly seeing the front and backside of the expanding 1612~MHz OH shell. In the case of WX~Psc the position of the redshifted maser was ambiguous, but we could not detect any trace of emission spatially coinciding with the blueshifted feature.

\begin{figure*}[htbp]
\centering
\includegraphics[width=0.8\textwidth, trim=0mm 10mm 0mm 10mm, clip]{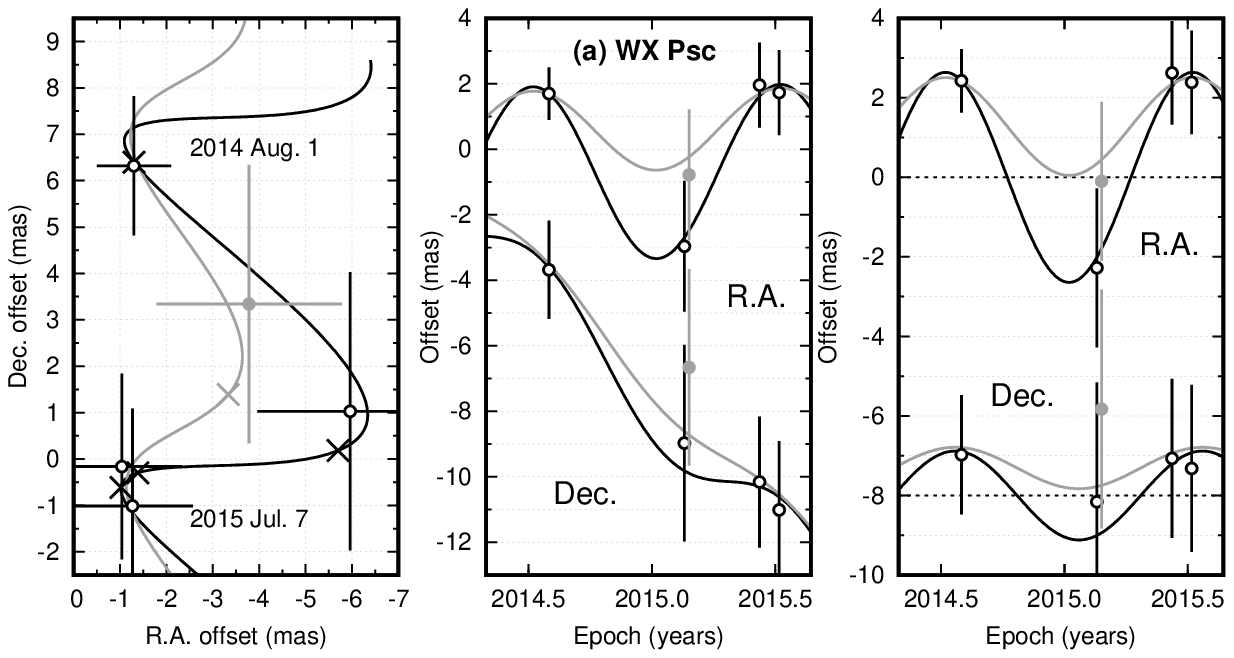}
\includegraphics[width=0.8\textwidth, trim=0mm 10mm 0mm 10mm, clip]{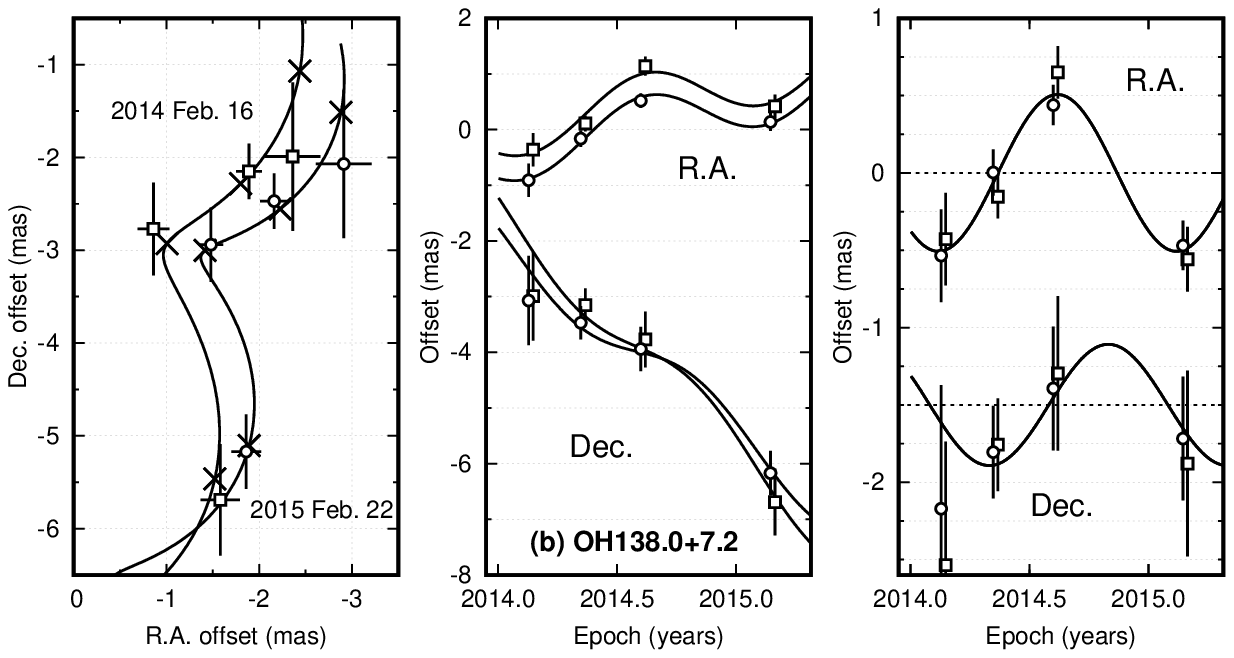}
\caption{
Parallax and proper motion fitting for 1612 MHz OH maser spots associated with (a) WX~Psc and (b) OH138. Left: Maser spot positions on the sky with respect to the shifted phase tracking centers; first and last epochs are labeled. The positions expected from the fitted models are indicated with crosses. Middle: Maser spot offsets in RA and Dec as a function of time, with the best parallax and proper motion fits. Right: Parallactic motions of maser spots in RA and Dec shown after removing the linear proper motions. In case of WX~Psc, two possible models are shown due to the ambiguity in the maser spot position in Epoch~II. For clarity, in the middle and right panels constant shifts are added to the offsets, with spots also shifted in time by a small amount in case they are in the same epoch (7 days).}
\label{fig:astrometry}
\end{figure*}

Maser motion is characterized by a combination of a linear proper motion component and a sinusoidal component from the annual parallax. Figure~\ref{fig:astrometry} shows the fitted proper motions and parallaxes of WX~Psc and OH138, while Table~\ref{table:astrometry} lists the measured and calculated values. In the case of WX~Psc, we could only use a single detected maser spot to derive the maser motions, and although the proper motion could be determined, we were only able to put an upper limit on the annual parallax (see below). For OH138, we could follow and fit the motion of two spots of the same maser feature, and the final proper motion and parallax values were derived using ``group fitting'' -- i.e. fitting both maser spots together by assuming a common distance. We checked our derived values by also fitting the parallax from the two spots individually, with all giving consistent results.

\begin{table*}[htbp]
\centering
\caption[]{\label{table:astrometry}Astrometric results of 1612 MHz OH masers around WX~Psc and OH138.}
\begin{tabular}{c @{\hskip 8 pt} c c c r r @{\hskip 0pt} c @{\hskip 0pt} c @{\hskip 0pt} c}
\tableline
\tableline
\noalign{\smallskip}
\multicolumn{1}{c}{Target} & Spot  & Epoch & V$_{\rm LSR}$ & \multicolumn{1}{c}{$\Delta\alpha$ cos $\delta$} & \multicolumn{1}{c}{$\Delta\delta$} & S$_{\rm 1.6GHz}$ & RMS & SNR\\
\noalign{\smallskip}
& ID & & (km s$^{-1})$ & \multicolumn{1}{c}{(mas)} & \multicolumn{1}{c}{(mas)} & \multicolumn{2}{r}{(mJy beam$^{-1})$}\\
\tableline
\noalign{\smallskip}
WX~Psc & \ding{109} & I    & $-$9.5 & $-$1.3 $\pm$ 0.8 &       6.3 $\pm$ 1.5 & 190 & 11.0 & 17\\
              &                     & IIa   & $-$9.3 & $-$6.0 $\pm$ 2.0 &       1.0 $\pm$ 3.0 &  54  & 7.4   & 7\\
              &                     & IIb   & $-$9.7 & $-$3.8 $\pm$ 2.0 &       3.3 $\pm$ 3.0 &  50  & 7.4    & 7\\
              &                     & III    & $-$9.4 & $-$1.0 $\pm$ 1.3 & $-$0.2 $\pm$ 2.0 & 104 & 14.4 & 7\\
              &                     & IV   & $-$9.5 & $-$1.3 $\pm$ 1.3 & $-$1.0 $\pm$ 2.1 & 128 & 13.0 &10\\
\noalign{\smallskip}
& (a)& \multicolumn{3}{c}{$\mu_{\alpha}$cos $\delta$ = 0.1 $\pm$ 0.4 mas yr$^{-1}$} & \multicolumn{2}{c}{$\mu_{\delta}$ = $-$7.5 $\pm$ 0.7 mas yr$^{-1}$} & \multicolumn{2}{c}{$\pi$ = 2.9 $\pm$ 0.8 mas}\\
& (b)& \multicolumn{3}{c}{$\mu_{\alpha}$cos $\delta$ = 0.1 $\pm$ 0.6 mas yr$^{-1}$} & \multicolumn{2}{c}{$\mu_{\delta}$ = $-$7.5 $\pm$ 1.0 mas yr$^{-1}$} & \multicolumn{2}{c}{$\pi$ = 1.3 $\pm$ 1.2 mas}\\
\tableline
\noalign{\smallskip}
OH138.0+7.2 & \ding{109} & I    & $-$47.0 & $-$2.9 $\pm$ 0.3 & $-$2.1 $\pm$ 0.8 & 472 & 7.9    &   60\\
              &                     & II   & $-$47.2 & $-$2.2 $\pm$ 0.15 & $-$2.5 $\pm$ 0.3 & 343 & 11.6 &   30\\
              &                     & III  & $-$47.0 & $-$1.5 $\pm$ 0.13 & $-$2.9 $\pm$ 0.4 & 570 & 8.8    &  65\\
              &                     & IV & $-$46.9 & $-$1.9 $\pm$ 0.16 & $-$5.2 $\pm$ 0.4 & 516 & 8.2    &  63\\
\noalign{\smallskip}
& & \multicolumn{3}{c}{$\mu_{\alpha}$cos $\delta$ = 0.92 $\pm$ 0.14 mas yr$^{-1}$} & \multicolumn{2}{c}{$\mu_{\delta}$ = $-$3.48 $\pm$ 0.31 mas yr$^{-1}$} & \multicolumn{2}{c}{$\pi$ = 0.47 $\pm$ 0.06 mas}\\
\cmidrule[0.04em](lr){2-9}
\noalign{\smallskip}
              & \ding{114}    & I    & $-$46.7 & $-$2.4 $\pm$ 0.3 & $-$2.0 $\pm$ 0.8 & 190 & 7.9     &  24\\
              &                     & II   & $-$46.9 & $-$1.9 $\pm$ 0.14 & $-$2.2 $\pm$ 0.3 & 408 & 11.6  &  35\\
              &                     & III  & $-$46.6 & $-$0.9 $\pm$ 0.17 & $-$2.8 $\pm$ 0.5 & 156 & 8.8    &  18\\
              &                     & IV & $-$46.6 & $-$1.6 $\pm$ 0.2 & $-$5.7 $\pm$ 0.6 & 130 & 8.2    &  16\\
\noalign{\smallskip}
& & \multicolumn{3}{c}{$\mu_{\alpha}$cos $\delta$ = 0.93 $\pm$ 0.27 mas yr$^{-1}$} & \multicolumn{2}{c}{$\mu_{\delta}$ = $-$4.33 $\pm$ 0.73 mas yr$^{-1}$} & \multicolumn{2}{c}{$\pi$ = 0.61 $\pm$ 0.12 mas}\\
\cmidrule[0.04em](lr){2-9}
\multicolumn{2}{r}{Group fitting} & \multicolumn{3}{c}{$\mu_{\alpha}$cos $\delta$ = 0.93 $\pm$ 0.21 mas yr$^{-1}$} & \multicolumn{2}{c}{$\mu_{\delta}$ = $-$3.89 $\pm$ 0.53 mas yr$^{-1}$} & \multicolumn{2}{c}{$\pi$ = 0.52 $\pm$ 0.09 mas}\\
\tableline
\noalign{\smallskip}
\end{tabular}\\
\vspace{-10 pt}
\justify
{\bf Notes.} Spot ID symbols are identical to ones used on Fig.~\ref{fig:astrometry}, with the exception of WX~Psc spot IIb plotted in gray. In the case of WX~Psc, (a) shows the model fit in black and uses spot IIa, while (b) shows the model fit in gray and uses spot IIb for Epoch~II with the rest of the epochs being identical between the two variations (see Fig.~\ref{fig:astrometry}). Positions are relative to the following J2000.0 phase centers: $(\alpha,\delta)_{\rm WX~Psc}$=(01$^{\rm h}$06$^{\rm m}$26$\fs$02574,$+$12$\degr$35$\arcmin$52$\farcs$8242) and $(\alpha,\delta)_{\rm OH138}$=(03$^{\rm h}$25$^{\rm m}$08$\fs$42975,$+$65$\degr$32$\arcmin$07$\farcs$0900). Position errors are the same as error bars shown on figures and are derived from subarray imaging (see ``triangle baseline errors'' in Sect.~\ref{subsec:triangle}). OH138 Spot \ding{114} errors were rescaled from Spot \ding{109} errors by the difference in SNR. RMS values refer to the average noise in line-free channels of the image cubes. 
\end{table*}

The proper motion values contain both the systemic proper motions related to the stellar movement and the internal motions of the OH masers relative to their host stars. It is not possible from our few detected maser spots to unambiguously separate these two components, however this does not change the value of the parallax as long as the masers are not accelerating. Also, as we are tracing the front side of an expanding shell in the line-of-sight of the star (see Fig.~\ref{fig:maps}), internal motions should not dominate the derived proper motions \citep[c.f.][]{vanlangevelde2000}.

In the case of WX~Psc, there is a large ambiguity in the derived parallax value from the difficulty of tracing the same maser component between epochs. In Epoch~II the brightest part of the feature separates into two spots in adjacent velocity channels ($-$9.3 and $-$9.7 km s$^{-1}$), both evenly shifted to the brightest spot's velocity of $-$9.5 km s$^{-1}$ in other epochs. We fitted the parallax model using both spots individually and while the proper motion values agree, there is a discrepancy in the amplitude of the parallaxes (see the black and gray models in Fig.~\ref{fig:astrometry} for the fits using the spot at $-$9.3 and $-$9.7 km s$^{-1}$ respectively). The parallax values also have a large uncertainty due to the timing of our observations, as the fitting errors are dominated by the measurement errors in Epoch~II. Therefore, we conservatively quote uncertainties that are double the formal fitting errors \citep[c.f.][]{reid2009}, since effectively our parallax and proper motion solutions have only one degree of freedom. Averaging the two channels provides a parallax value between the two extremes with all options resulting in $\chi^2$$<$1 solutions. At this time, it is impossible to better constrain the parallax value of WX~Psc and we only quote a 3$\sigma$ upper limit of $\pi$$\lesssim$5.3~mas as the final result.

We also compared the fitted parallax values for OH138 using unflagged data and data that were flagged as described in Sect.~\ref{subsec:triangle}. All measurements were derived from the same data reduction and were assigned identical astrometric uncertainties for comparison. When data are flagged, we get the results shown on Fig.~\ref{fig:astrometry} and listed in Table~\ref{table:astrometry}: $\pi$=0.52$\pm$0.09 mas with $\chi^2$=4.2. As the result is derived using group fitting, the quoted uncertainty is the formal fitting error multiplied by $\sqrt{2}$ to allow for the possibility of correlated positions for the two maser spots \citep[c.f.][]{reid2009}. For the unflagged data the fitted values are: $\pi$=0.36$\pm$0.14 mas with $\chi^2$=10. We can see a reduction in both the $\chi^2$ values and the $\sigma$ fitting errors in the flagged data in comparison to the unflagged data, though the two parallaxes are consistent within their uncertainties.  The error in the parallax obtained with the flagged data is lower and that reflects the additional effort made in mitigating systematic errors, thus we conclude $\pi$=0.52$\pm$0.09 mas as the parallax estimate for OH138.

On a side note, we would like to mention the limitations of a common practice in parallax papers. Astrometric errors are often calculated by adjusting the error values in each coordinate independently to attain a $\chi_{\rm red}^{2}$=$\chi^{2}/K$$\sim$1 value in the proper motion and parallax fitting, where $K$ is the number of degrees of freedom in the model \citep[e.g.][]{reid2009,imai2012}. This might be a valid method for estimating errors when fitting highly overdetermined models with many epochs and maser spots, but it has two problems: $K$ is not always simply ``data points'' minus ``model parameters'' and the $\chi^{2}$ value also has an uncertainty due to measurement errors \citep{andrae2010}. In our case, $K$ is essentially 1 and 3 for the fitting of WX~Psc and OH138 respectively, thus e.g. we can derive a $\chi_{\rm red}^{2}$$\approx$1.4 for the OH138 parallax. Although $\chi_{\rm red}^{2}$ values around unity are reassuring in evaluating the quality of the fit, the mentioned problems in calculating $\chi_{\rm red}^{2}$ are most severe when $K$ is small and renders $\chi_{\rm red}^{2}$ based error estimation and parallax fitting unreliable in our case.

\subsection{Comparison with Published OH Astrometry}
\label{subsec:oldOHastro}

Prior to our results, only 5 AGB stars had measured parallaxes based on astrometry of OH masers \citep{vanlangevelde2000, vlemmings2003, vlemmings2007}. These measurements used 1665~MHz and 1667~MHz main-line transitions and a variation of conventional source-switching and in-beam techniques, depending on the available calibrators for each source. In the case of conventional phase referencing (U~Her, R~Cas and W Hya), the total astrometric uncertainties were between 0.7--3~mas per epoch, depending on the target--calibrator separations and ionospheric conditions. Proper motion errors for RA and Dec were in the range of 0.3--3 mas yr$^{-1}$ and 0.3--5 mas yr$^{-1}$, respectively. The precision of the parallax estimates were 0.3--4~mas, where fitting errors depend on the number of observed epochs and maser spots used, although more spots do not necessarily imply a higher precision or accuracy. The astrometric uncertainties generally decreased when in-beam calibration was possible (S~CrB and RR~Aql, with calibrator separations of 20$^{\prime}$$-$24$^{\prime}$), to 0.2--1.5~mas per epoch. Fitting errors also decreased to 0.2--0.7 mas yr$^{-1}$, 0.3--0.7 mas~yr$^{-1}$ and 0.17--0.5~mas for the proper motions in RA, Dec, and the parallax, due to the smaller source separations and simultaneous observations of target--calibrator pairs.

From this, we can conclude that our astrometric precision per epoch is essentially the same as prior in-beam results indicating no reduction in error contributions of dynamic nature. On the other hand, the uncertainties in our parallax and proper motion fits of OH138 agree with previous in-beam investigations in spite of using fewer epochs (4 epochs per star compared to 5--17 epochs in previous works). The errors in the proper motion of WX~Psc also agree with this trend, while the parallax results were shown to be too ambiguous for a useful comparison. This might indicate that our flagging scheme explained in Sect.~\ref{subsec:triangle} works well in reducing static errors and can be of good use to multi-epoch maser VLBI investigations at low frequencies. However, these error metrics have to be treated with care as we have very few data points \citep[see][]{andrae2010}. Smaller errors can also be attributed to better sampling of the parallactic motion \citep[i.e. measuring at the RA extrema as opposed to a uniform sampling,][]{reid2009}, although our measurement spacings were also not optimal.

\subsection{Comparison with the Phase-Lag Technique}
\label{subsec:phaselag}

In the following section, we compare our derived trigonometric distances to distances derived with the ``phase-lag'' technique \citep[][hereafter called as ``phase-lag distances'']{vanlangevelde1990}, to try to evaluate their accuracy as distance measurement tools to evolved stars. As discussed in Sect.~\ref{sec:intro}, the 1612~MHz OH masers around OH/IR stars originate in saturated circumstellar shells, expanding outward at terminal velocity. As a result 1612~MHz OH masers are radially amplified and their intensity follow the underlying stellar pulsations.

Phase-lag distance measurements rely on monitoring these variations in the red- and blueshifted maser peaks to determine the $\tau_0$ time shift between the emission coming from the front and backside of the 1612 MHz OH maser shell \citep{vanlangevelde1990}. The $D$ linear diameter of the shell can be obtained using $D$=$c\tau_{0}$, where $c$ is the speed of light in vacuum. By also measuring the $\phi$ angular size of the OH shell through interferometric observations \citep{etoka2014}, we can get the $d_{pl}$ phase-lag distance to the star as $d_{pl}$=$D\phi^{-1}$=$c\tau_{0}\phi^{-1}$.

Table~\ref{table:phaselag} lists the two different distance estimates for our observed OH/IR stars, where the phase-lag values\footnote{Based on an on-going single-dish monitoring program with NRT, and imaging campaigns with the NRAO Karl G. Jansky Very Large Array (VLA) and the UK based Multi-Element Remotely Linked Interferometer Network (eMERLIN). For updates in the phase-lag distance project, refer to www.hs.uni-hamburg.de/nrt-monitoring.} are taken from \citet{engels2015b}. Trigonometric distances are the reciprocals of the measured parallax values described in Sect.~\ref{subsec:parallax}. In case of low significance results, prior assumptions can heavily influence the converted distance \citep{bailerjones2015}, such as the spatial distribution of the Galactic AGB population. However, as known distributions are poorly constrained and depend largely on the initial mass, we do not consider their effect on our distances at this time. In the case of WX~Psc, we quote only a lower limit of $\gtrsim$190~pc on the distance at a 3$\sigma$ significance level. For OH138, the measured distance is $\sim$$1.9$~kpc at a $>$5$\sigma$ significance.

Phase-lag distances agree with the trigonometric distances within the errors, but only OH138 can yield a proper comparison at this time. The phase-lag distance of OH138 gives a larger value by $\sim$15\%, which might be explained by the limited sensitivity of the interferometric observations in determining the angular size of the OH shell close to the systemic velocity where the 1612~MHz OH maser emission is the weakest. This can lead to an overestimation of the distance, although poor map quality also makes measurements more uncertain so we cannot consider this a systematic bias. Furthermore, the measured angular size also depends on the assumed OH maser shell model: deviations such as spherical asymmetry, non-isotropic radiation, or a thicker OH shell cause systematic errors and skew the measured distances, unless corrected by proper geometrical modeling \citep{etoka2010}.

\begin{table}[htbp]
\caption[]{\label{table:phaselag}Comparison of distance estimates.}
\begin{tabular}{l c c}
\tableline
\tableline
\noalign{\smallskip}
Method & WX~Psc & OH138.0+7.2\\
\noalign{\smallskip}
& distance (kpc) & distance (kpc)\\
\tableline
\noalign{\smallskip}
Trigonometric & $\gtrsim$0.2 & 1.9 $^{+0.4}_{-0.3}$\\
\noalign{\smallskip}
Phase--lag & 0.5 $\pm$ 0.1 & 2.2 $\pm$ 0.5\\
\noalign{\smallskip}
\tableline
\noalign{\smallskip}
\end{tabular}\\
\vspace{-10 pt}
\justify
{\bf Notes.} Trigonometric errors are based on the parallax fitting, with only a 3$\sigma$ lower limit quoted for WX~Psc. Phase-lags assume $\pm$20\% relative errors based on \citet{engels2015b}. 
\end{table}

\section{Summary and Outlook}
\label{sec:conclusions}

Over the last decade the field of OH maser astrometry has been in abeyance. In this paper, we used VLBA multi-epoch measurements to derive astrometric solutions of 1612 MHz OH masers around WX~Psc and OH~138.0+7.2, the first results since \citet{vlemmings2007}. We have shown that the largest limitation in attaining these low-frequency astrometric solutions is from direction-dependent static ionospheric errors that can cause large systematic shifts in the measured maser positions, and are very sensitive to the target--calibrator separation. This can cause serious problems even with in-beam phase referencing calibration techniques, and limit the accuracy of OH maser parallax measurements at 1.6~GHz. We tried to partially address this problem by introducing a new method to identify data affected by systematic errors. It is based on making many different three-antenna subarray images of the same compact emission and analyzing the spatial distribution of the peak positions between the various images. We also showed that it is possible to get a good estimate of astrometric uncertainties using this technique, although variations in antenna number should be explored to increase the robustness of the results. It would also be prudent to further investigate the correlation between the static ionospheric errors of different antennas in VLBI arrays.

Based on our corrected OH maser astrometry, the resulting annual parallax of OH138 is $\pi$=0.52$\pm${0.09}~mas, making it the first sub-mas OH maser parallax measured. This corresponds to a distance of $d_{\pi}$=1.9$^{+0.4}_{-0.3}$~kpc, which places the source in the thick disk region $\sim$240~pc above the Galactic plane. For the parallax of WX~Psc only a 3$\sigma$ upper limit of $\pi$$\lesssim$5.3~mas could be determined, placing it at a distance of $d_{\pi}$$\gtrsim$190~pc. The present results are also the first trigonometric distance measurements to enshrouded OH/IR stars, with long period variability (650 and 1410 days) and high mass loss rates of $\sim$10$^{-5}$~M$_\odot$~yr$^{-1}$. Finally, we used our trigonometric distances to evaluate the phase-lag distance estimates of these sources and found the two independent methods to be in good agreement with each other. Due to the large uncertainty in the derived distances, it is necessary to continue the parallax observations for a more comprehensive analysis. This would also make it possible to explore additional error estimates, e.g. cross-validation or bootstrapping \citep{andrae2010}, to better confine the errors of our astrometry and evaluate the newly introduced triangle baseline errors.

Working out the technical issues of low-frequency OH maser astrometry will make it possible to accurately and precisely measure the trigonometric distances to a large number of OH/IR stars and study potential questions in late stellar evolution. One main scientific topic is the study of the period--luminosity relationship of Galactic long period variable AGB stars and the evolutionary connection between Miras and OH/IR stars. Accurate distances can also help us better understand, e.g. how metallicity or mass loss influence stellar pulsation and evolution, and they can also help us put constraints on AGB evolutionary models. Additionally, stellar OH masers can open up possibilities as new relaxed tracers of Galactic dynamics, and by measuring their distance we could study the motion of material in the inter-arm regions and thick disk. We will present case-studies for these topics using our newly derived OH/IR distances in a follow-up paper.

As a technical outlook for high-precision low-frequency astrometry, we are also developing and testing the {\it MultiView} technique mentioned in Sect.~\ref{sec:intro} as an alternative approach to in-beam phase-referencing calibration. MultiView is based on the 2D modeling of the phase screen around the target by using scans on multiple calibrators. This makes it possible to fully compensate for the spatial static component of the ionosphere, even with calibrators several degrees away. We demonstrated this technique in \citet{rioja2017}, achieving complete atmospheric mitigation and effectively reaching the thermal noise limit in our measurements. We plan to further develop and test these techniques to support our mentioned science goals.

Finally, we note that the Square Kilometer Array (SKA) is planned to be an important part of future VLBI networks \citep{paragi2015,green2015,imai2016}. As this will concentrate on the ionospheric dominated low frequencies, it is important to understand and solve the limiting factors of OH astrometry and parallax measurements, which will allow us to take full advantage of the astrometric capabilities of the SKA. 

\acknowledgments
We would like to thank Dieter Engels and Willem Baan for the helpful and interesting discussions related to AGB stars and maser theory.
We also give our gratitude to the referee for providing a very thorough review and whose suggestions led to many critical improvements in this work.
GO and RB would like to acknowledge the Monbukagakusho scholarship of the Ministry of Education, Culture, Science and Technology (MEXT), Japan for financial support. GO and HI were supported by the JSPS Bilateral Collaboration Program and KAKENHI programs 25610043 and 16H02167. 
GO, RD, MR and HI acknowledge support from the DFAT grant AJF-124. DT was supported by the ERC consolidator grant 614264.
We acknowledge the use of the Astrogeo Center Database of brightness distributions, correlated flux densities, and images of compact radio sources produced with VLBI.

{\it Facilities:} \facility{VLBA}.

\end{document}